%% file: nuprimerice.tex
\documentclass[12pt]{article}
\usepackage{makeidx}
\usepackage{epsfig}
\usepackage{wrapfig}
\usepackage{epic}
\usepackage{hyperref}
\usepackage{float}
\usepackage{amsmath}
\usepackage{amssymb}
\usepackage{fancyheadings}
\usepackage{setspace}
\input macpi.tex

\newtheorem{definition}{Definition}[section]

\makeindex

\begin{document}
\title{ 
        {\Huge A Mathematical Primer on Water Ice}  \\
        \vspace{0.01in} \rule{150mm}{0.1mm} \\ \vspace{0.1in}
        {\LARGE L. Ridgway Scott} \\ \vspace{0.0in}
        {\Large University of Chicago} \\ \vspace{0.01in}
}
\author{}
 \def\thepage{}\maketitle\pagenumbering{arabic}
\tableofcontents
\newpage
\input iceintro.tex

\input iceoneall.tex

\input icetwo.tex
\bibliography{dhd}
\bibliographystyle{plain}
\end{document}

%% file: macpi.tex
\newcommand{\Reyuls}{\mathbb{R}}
\newcommand{\Integrs}{\mathbb{Z}}

\newcommand{\parp}{{{\rm I}{\relax\ifmmode\mskip-\thinmuskip\relax\else\kern-.22em\fi}{\bf P}}}

\newcommand{\Pfortran}{{\parp}{\kern-.03em{\rm fortran}}}
\newcommand{\pfbf}{{\parp}{\kern-.03em{fortran}}}
\newcommand{\PC}{{\parp}{\kern-.03em{\rm C}}}

\newcommand{\Planguages}{{\parp}{\kern-.05em{\rm languages}}}
\newcommand{\half}{{\textstyle\frac{1}{2}}}

\newcommand{\norm}[1]{\Vert #1 \Vert}

\newcommand{\Forall}{\quad\forall}

\newcommand{\set}[2]{\left\lbrace #1 \; \big| \; #2 \right\rbrace}
\newcommand{\bigset}[2]{\bigg\lbrace #1 \; \big| \; #2 \bigg\rbrace}

\floatstyle{ruled}
\newfloat{email}{ht}{aux}[section]
\floatname{email}{E-mail}

\floatstyle{boxed}
\newfloat{code}{ht}{aux}[section]
\floatname{code}{Program}

\newcommand{\defdex}[1]{{\bf #1}\label{`#1'}\index{#1}} 

\newcommand{\sexshun}[1]{Section~\ref{#1}}

\newcommand{\figgar}[1]{Figure~\ref{#1}}

\newcommand{\defnn}[1]{Definition~\ref{#1}}

\newcommand{\eqnn}[1]{(\ref{#1})}

\newcommand{\Rtre}{{\Reyuls}^3}
\newcommand{\Ztre}{{\Integrs}^3}

\newcommand{\ut}[1]{\rlap{${}{\lower .3ex\hbox{$_{_\sim}$}}$}{#1}}

\newcommand{\chooze}[2]{\begin{pmatrix} #2\\#1\end{pmatrix}}

\newcommand{\antiparallel}{\not\kern-0.15em\Vert}

%% file: iceintro.tex
The following is an update of an earlier paper \cite{lrsBIBjh}.
The original version of \cite{lrsBIBjh} was written a decade earlier
and circulated informally.
That version was cited several times in 2012 and 2013.
The publication \cite{lrsBIBjh} in 2021 was done so that it could be
formally referenced in \cite{lrsBIBjc}, and it involved only slight modifications.
But much has transpired in the last decade that requires a significant revision.
Rather than calling this the second edition of \cite{lrsBIBjh}, we chose a slightly
longer, and more precise, name.

Water adopts many different crystal structures in its solid form
\cite{fletcher2009chemical,hobbsice,petrenko1999physics}.
These provide insight into potential structures of water even in its 
liquid phase \cite{ref:fasediagram2ndshell,ref:revuwaterstruk},
and they can be used to calibrate pair potentials used for simulation of water 
\cite{ref:SRicetestpotentice,ref:whaticecanteachwatermodel,ref:meltpointwatermodels}.
In crowded biological environments, water may behave more like ice than bulk water.
The different ice structures have different dielectric properties
\cite{ref:iceIItoVIdielectriconst}.
This brief primer is intended to facilitate further research.

The human-scale morphology of ice is legendary, with multiple words to describe
forms of snow \cite{eira2013traditional},
with the beauty of individual snow crystals \cite{ref:snowcrystalphysics}, 
and with other bizarre forms in nature such as hair ice \cite{ref:hairicevidence}.
Crystalline ice is necessarily elastic, but in nature ice exhibits
plasticity \cite{ref:icephasediagramTip4P,ref:wildiceplasticity} as well.
However, all of this morphology depends on the molecular structure of water ice 
at a much smaller scale that is our focus here.

\begin{wrapfigure}{r}{2.7in}
\centerline{\includegraphics[width=2.5in]{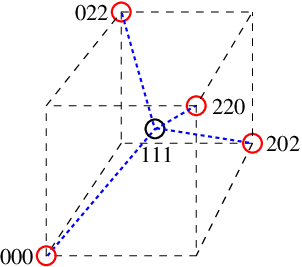}}
\caption{The tetrahedral basis for the crystal structures of both forms of ice I.
The triples of numbers are the Cartesian coordinates of the water positions in
a cube of side two units.
This local structure forms the basis for both the cubic structure of the diamond 
lattice for ice Ic and the hexagonal structure of ice Ih.}
\label{fig:onecube}
\end{wrapfigure}

\section{Introduction to ice}
\label{chap:iceintro}

There are many crystal structures of ice that are topologically tetrahedral
\cite{ref:tetrahedralorderparam},
that is, each water molecule makes four hydrogen bonds with other water molecules,
even though the basic structure of water is trigonal \cite{ref:finneywatermolecule}.
Two of these crystal structures (Ih and Ic) are based on the same exact 
local tetrahedral structure, as shown in \figgar{fig:onecube}.
Thus a subtle understanding of structure is required to differentiate them.

We refer to the tetrahedral structure depicted in \figgar{fig:onecube} as
an exact tetrahedral structure.
In this case, one water molecule is in the center of a square cube (of side length
two), and it is hydrogen bonded to four water molecules at four corners of the cube.
The triples of numbers represent the Cartesian coordinates of the various water 
locations.
The distance between oxygen centers in these coordinates is $\sqrt{3}$ for the
waters that are directly hydrogen bonded, as indicated by the dotted lines in
\figgar{fig:onecube}, and the distance between the next-nearest neighbors is
$2\sqrt{2}$.
The actual distance for ice Ih or Ic is believed \cite{ref:paulingiceoneh}
to be 2.76 {\AA}ngstroms, so to convert our coordinates to {\AA}ngstroms 
we must multiply by $L= 1.5935$ {\AA}ngstroms.

This value of $L$ can be checked using the known density of ice.
We will see that the density of ice Ih and Ic in our units is one water per
eight cubic units, which corresponds to about one water per 32.37 cubic {\AA}ngstroms
if we use \cite{ref:paulingiceoneh}.
The atomic weight of H$_2$O is 18.01528 amu.
One amu is $\approx 1.660539 \times 10^{-24}$ grams, so one H$_2$O 
weighs $\approx 2.9915  \times 10^{-23}$ grams.
So water ice has a density of
$$
\frac{2.9915  \times 10^{-23}}{8 L^3 \times 10^{-24} }\;\hbox{grams per cubic centimeter}
$$
The maximum density of water occurs at 4 degrees Centigrade, 
and at this temperature one gram of water occupies one cubic centimeter, or one milliliter.
But when water freezes, at 0 degrees Centigrade, its density is only 0.9167 grams per cubic 
centimeter \cite{harvey2012properties}.
This suggests $L\approx 1.598$, which is consistent with the value based on the
data in \cite{ref:paulingiceoneh} to the number of digits given.
This would imply that the distance between two oxygens in ice I is 2.767 {\AA}ngstroms
at 0 degrees Centigrade.
The density of ice increases as temperature decreases, so at $-40$ degrees (Centigrade or
Fahrenheit), the distance would decrease to 2.762 {\AA}ngstroms.

The ice structures are not simple repeats of the basic cube in Figure \ref{fig:onecube}.
Instead, this unit is utilized in different ways to form different forms of ice I.
In other forms of ice, this tetrahedral structure is distorted, but the four hydrogen
bonds remain.

Notice that we have not specified the location of the hydrogen (donors) for the
hydrogen bonds in \figgar{fig:onecube}.
We use dotted lines to indicate the hydrogen bonds inside the cube,
independent of where the hydrogens are located.
In general, there are six different specifications for the hydrogens for the 
central water molecule located at the center of the box, since there are four
possible locations for each pair of hydrogens:
$$
6=\chooze{2}{4}.
$$
For each of these six pairings, there is a compatible 
assignment of hydrogens for the four corner water molecules (each edge must
contain precisely one hydrogen).
However, when one continues this process to their neighbors, constraints on the
hydrogen positions become significant.
In some forms of ice, the positions of the hydrogens are fixed, and we can indicate
the directions of the hydrogen bonds by arrows (cf.~\figgar{fig:icetwosidevu}).

One way to think of the structure of ice is as a graph with a type of periodic
structure.
In this way, there is a natural finite graph that generates this periodic structure
\cite{ref:K4crystalattice}.
These graphs differentiate the different forms of ice, and they help to visualize
the structure.
Let us introduce these notions more precisely.

\subsection{Lattices in $\Rtre$}
\label{sec:lattre}

A lattice \cite{dubrovin} consists of a set of points generated by shifting the
origin in three directions repeatedly.
A lattice is specified by three vectors; examples are given below
in \eqnn{eqn:iceIcgeneratrs} and \eqnn{eqn:iceIsgeners}.

\begin{definition}
\label{def:lattice}
Suppose that $g^1,g^2,g^3\in \Rtre$ are not co-planar.
A \defdex{lattice} ${\cal L}(g^1,g^2,g^3)$ in $\Rtre$ is the set of points
$$
{\cal L}(g^1,g^2,g^3)=\bigset{x_\alpha=\sum_{i=1}^3 \alpha_i g^i}{\alpha\in\Ztre},
$$
where $g^1,g^2,g^3\in\Rtre$ are called the generators of the lattice.
\end{definition}

Thus a lattice consists of the vertices of a tiling of $\Rtre$ obtained 
with simple boxes with possibly slanted sides.
Lattices make sense in any number of dimensions, with the number of generators
being the dimension.
The vertices of the parallelograms in \figgar{fig:symmetriceoneh} provide an 
example of a lattice in two dimensions.
Since we are primarily interested in three-dimensional lattices, we have simplified
to this case.

A lattice forms a group under the addition
\begin{equation}
\label{eqn:crystalgroup}
x_\alpha+x_\beta= \sum_{i=1}^3 (\alpha_i+\beta_i) g^i\Forall \alpha,\beta\in\Ztre,
\end{equation}
and this group is isomorphic to $\Ztre$.
We can denote this isomorphism by $\phi$, where
\begin{equation}
\label{eqn:groupiso}
\phi(\beta)=x_\beta \Forall \beta\in\Ztre.
\end{equation}

The \defdex{fundamental domain} of a lattice ${\cal L}(g^1,g^2,g^3)$ is the set
\begin{equation}
\label{eqn:fundomlat}
\Omega_{\cal L}=\set{y=x_1 g^1+x_2 g^2+x_3 g^3}{x\in\Rtre,\; 0\leq x_i\leq 1,\; i=1,2,3}.
\end{equation}
The shapes of the fundamental domain of a lattice are crucial in determining 
certain symmetries of crystals, and they are useful in visualizing the
repeating pattern of crystals.
However, these symmetries are primarily of interest when determining the vertex 
positions based on X-ray data.

\subsection{Crystals in $\Rtre$}
\label{sec:crystalatis}

A crystal is more complex than a lattice.
It has a set of points ${\cal P}\subset\Rtre$ that form the basis for the crystal
unit, and these are shifted throughout space using a lattice.
Typically, ${\cal P}\subset\Rtre$ is a finite set.

\begin{definition}
\label{def:indepcrys}
We say that a set ${\cal P}\subset\Rtre$ is non-degenerate with respect to a
lattice ${\cal L}(g^1,g^2,g^3)$ if for any $p,q\in {\cal P}$ and 
$x_\alpha\in {\cal L}(g^1,g^2,g^3)$, $p=q+x_\alpha$ implies $x_\alpha=0$.
\end{definition}

\begin{definition}
\label{def:crystal}
Suppose $g^1,g^2,g^3\in\Rtre$ are the generators of a lattice ${\cal L}(g^1,g^2,g^3)$,
and ${\cal P}\subset\Rtre$ is non-degenerate with respect to ${\cal L}(g^1,g^2,g^3)$.
A \defdex{crystal} ${\cal C}$ in $\Rtre$ is a graph ${\cal C}=({\cal V},{\cal E})$
whose vertices ${\cal V}$ comprise a set of points
\begin{equation}
\label{eqn:crystalset}
{\cal V}=\set{x_\alpha+p}{p\in{\cal P},\; x_\alpha\in{\cal L}(g^1,g^2,g^3)},
\end{equation}
and whose edges ${\cal E}$ are also generated by a finite set 
${\cal S}\subset{\cal V}\times{\cal V}$ of edges:
\begin{equation}
\label{eqn:crystaledge}
{\cal E}=\set{(x_\alpha+v^1,x_\alpha+v^2)}{(v^1,v^2)\in{\cal S},\; 
               x_\alpha\in{\cal L}(g^1,g^2,g^3)}.
\end{equation}
\end{definition}

One simple crystal ${\cal C}(g^1,g^2,g^3)$ is based on a lattice ${\cal L}(g^1,g^2,g^3)$, 
with ${\cal P}=\lbrace (0,0,0)\in\Rtre \rbrace$ and ${\cal S}=\set{(0,g^i)}{i=1,2,3}$.
Notice that the set ${\cal S}$ is not a set of edges based on points in ${\cal P}$.
That is, the pair $({\cal P},{\cal S})$ does not form a graph itself.

\section{Comparing crystals}
\label{sec:comparystal}

Ice takes on various crystal forms under different temperature and pressure
conditions.
We would like to have ways to understand the structures that compare, contrast,
and differentiate them.
We review three different concepts that distinguish between common ice forms.
We will see that some forms of ice (e.g., Ih and Ic) appear superficially similar, and
this motivates us to look into more complex measures that distinguish them.

\begin{wrapfigure}{r}{2.4in}
\vskip -60pt
\begin{center}
\begin{tabular}[t]{|c|c|c|c|}
\hline
distance &\AA & $N_h$ & $N_c$ \\
\hline
    1.7321&  2.76 &  4 &  4 \\
    2.8284&  4.51 & 12 & 12 \\
    2.8868&  4.60 &  1 & \\
    3.3166&  5.29 &  9 & 12 \\
    4.0000&  6.37 &  6 &  6 \\
    4.0415&  6.44 &  6 & \\
    4.3589&  6.95 &  9 & 12 \\
    4.6188&  7.36 &  2 & \\
    4.8990&  7.81 & 18 & 24 \\
    5.1962&  8.28 &  9 & 16 \\
    5.4160&  8.63 & 12 & \\
    5.4467&  8.68 &  3 & \\
    5.6569&  9.01 &  6 & 12 \\
    5.6862&  9.06 &  6 & \\
    5.9161&  9.43 & 18 & 24 \\
\hline
    total &     &  121 &122 \\
\hline
\end{tabular}
\end{center}
\vskip -10pt
\caption{Radial distances between oxygen centers in two different forms of
ice one: ice Ih and ice Ic.
The left two columns are the radial distances in mathematical and physical
units, and the third and fourth columns are the
numbers of water molecules at that distance: $N_h$ for ice Ih and $N_c$ for ice Ic.
}
\label{tabl:radisticeo}
\end{wrapfigure}

\subsection{Radial distribution function}
\label{sec:radistfuncn}

The \defdex{radial distribution function} is one measure often used to describe crystals.
The primary one in the context of ice is the oxygen-oxygen radial distribution function.
This is a type of probability distribution function related to the likelihood of
finding a neighboring oxygen at a given distance $r$.

For true crystals, such a distribution is discrete, that is, a sum of delta functions
at distinct distances, due to the fact that the positions of the atoms in 
\eqnn{eqn:crystalset} have a periodic structure.
Moreover, all crystals based on the local tetrahedral structure in \figgar{fig:onecube}
have the property that the first (smallest $r$) nonzero part of the radial distribution 
is the same, corresponding to the fact that each oxygen has four neighbors at a
distance of $\sqrt{3}$ in the coordinates of \figgar{fig:onecube}.
The second nonzero part of the radial distribution
is also the same, corresponding to the fact that the (twelve) next nearest neighbors 
are at the opposite corners of a cube of side two, and thus are all at a 
distance of $2\sqrt{2}$ in the coordinates of \figgar{fig:onecube}.
The topology of the hydrogen bond connections for these nearest neighbors for ice I
are depicted in \figgar{fig:icelocalogy}.

\begin{figure}
\centerline{(a)\includegraphics[width=2.3in,angle=-90]{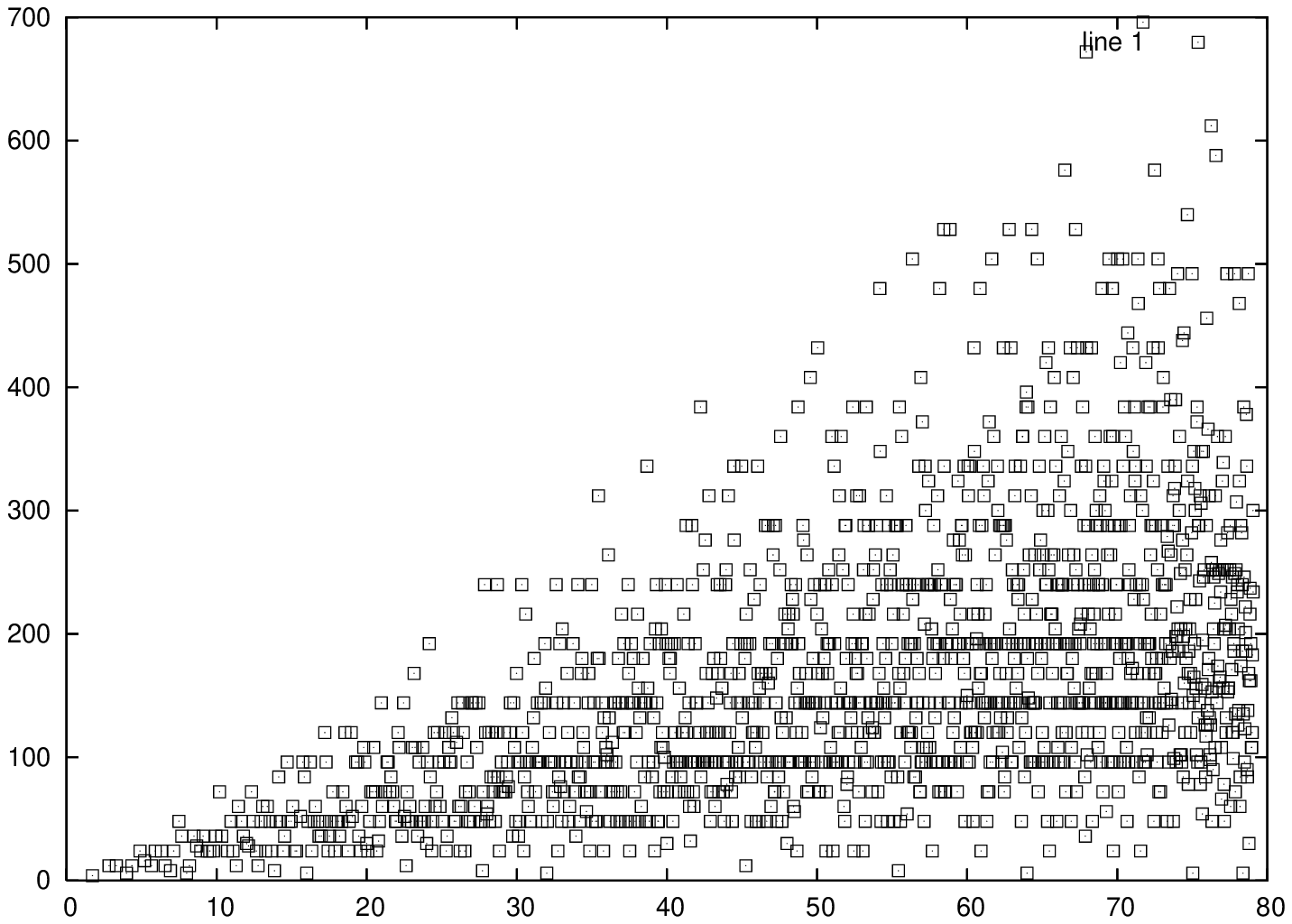}
            (b)\includegraphics[width=2.3in,angle=-90]{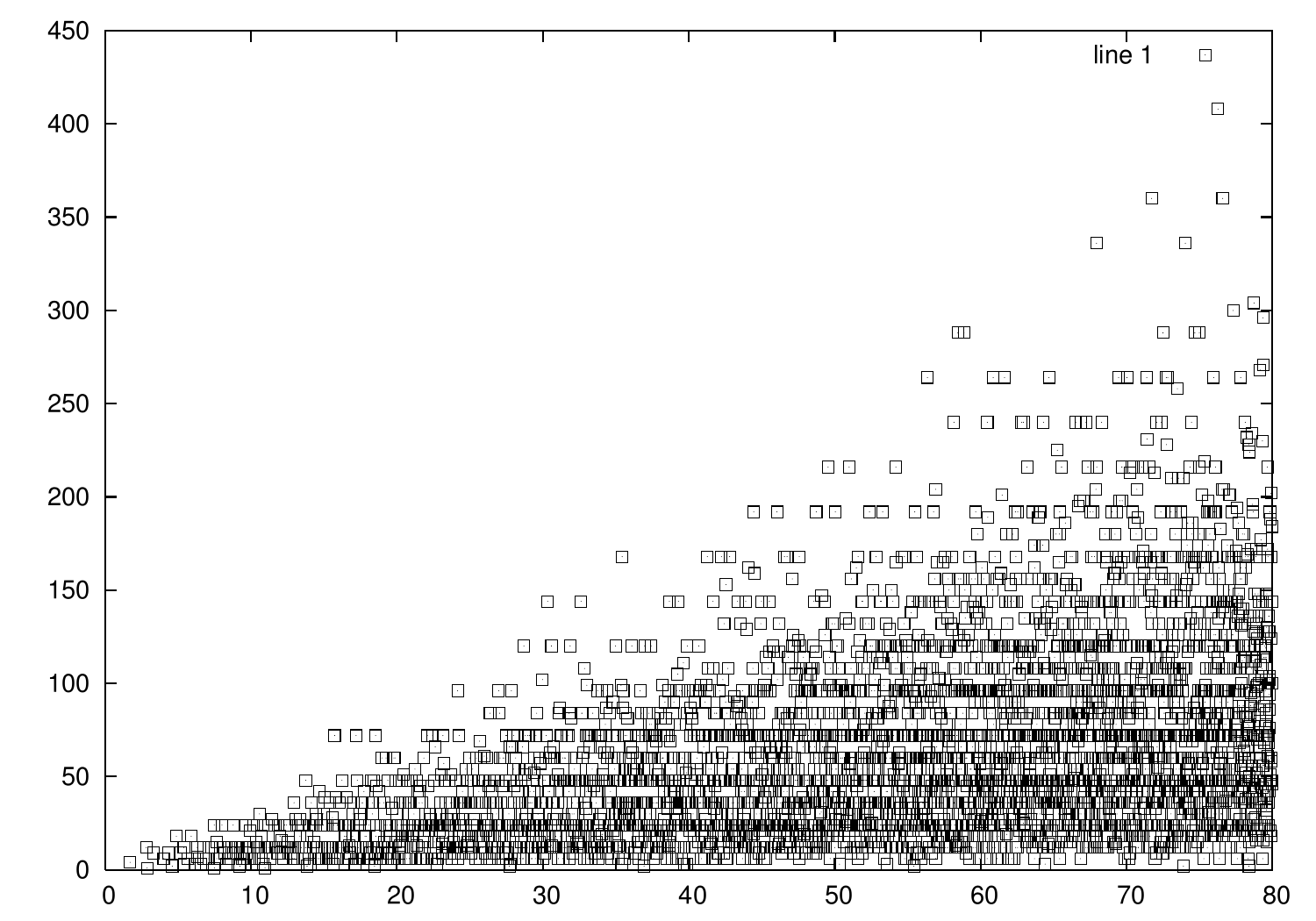}}
\caption{The radial distribution of ice Ic (a) and ice Ih (b).
The horizontal axis is in mathematical units for ice, as in the first column
in the table in Figure \ref{tabl:radisticeo} (multiply by 1.5935 to get {\AA}ngstroms
or by $0.8433$ to get atomic units).
The vertical axis represents the number of oxygen pairs at the given distance, that is,
the third and fourth columns in the table in Figure \ref{tabl:radisticeo}.
}
\label{fig:diamondrdist}
\end{figure}

There is a significant difference between the radial distribution function and
the radial density function.
The latter is the former divided by $4\pi r^2$.
Thus the first peak for tetrahedral water would have a height of $1/3\pi$ and
the second peak would have a height of $3/8\pi$.
Although the first two points of the radial functions for ice Ih and Ic are the
same (see \figgar{tabl:radisticeo}), their overall distributions are quite different
as shown in Figure \ref{fig:diamondrdist}.
Ice Ic has a sparser set of distribution points than Ih as illustrated in
\figgar{tabl:radisticeo}.
This pattern continues for larger distances; there are 96 distances in Ic 
less than twenty, but 239 distances in Ih less than twenty.
(Both crystals have thirty molecules at a distance of twenty.)
Because of the different radial densities, ice Ih and Ic have different energies
\cite{ref:bjerrumicerevuscience}.
Ice Ih has the lower energy.

\subsection{Quotient graph}
\label{sec:quotiegraph}

For any crystal ${\cal C}=({\cal V},{\cal E})$ generated from a lattice
${\cal L}(g^1,g^2,g^3)$, the natural lattice group defined via \eqnn{eqn:crystalgroup} 
acts on ${\cal C}$ in a natural way, and it leaves the crystal invariant:
\begin{equation}
\label{eqn:crystalinvar}
\begin{split}
&x_\beta (x_\alpha+p)= x_\alpha +x_\beta+p=x_{\alpha +\beta}+p,\\
&x_\beta (v^1,v^2)=(x_\beta + v^1,x_\beta + v^2),
\end{split}
\end{equation}
for all $\beta\in\Ztre$.
Note that we can think equivalently of \eqnn{eqn:crystalinvar} defining the 
action of ${\cal L}(g^1,g^2,g^3)$ on ${\cal C}$, or the 
action of $\Ztre$ on ${\cal C}$, where we make the obvious identification. 
Thus it is natural to consider the quotient graph ${\cal C}/{\cal L}(g^1,g^2,g^3)$
with respect to this group action.
Let us define this quotient graph.

The vertices of this quotient graph are equivalence classes of vertices in ${\cal V}$.
We say two points $v^1,v^2\in{\cal V}$ are equivalent with respect to the
group ${\cal L}(g^1,g^2,g^3)$ if, for some $\beta\in\Ztre$,
\begin{equation}
\label{eqn:vertxeqiv}
v^1= v^2 +x_\beta.
\end{equation}
We define ${\cal V}/{\cal L}(g^1,g^2,g^3)$ to be the set of equivalence classes
with respect to the equivalence relation \eqnn{eqn:vertxeqiv}.
Each $v^i=p^i+x_{\alpha_i}$, so \eqnn{eqn:vertxeqiv} means that 
\begin{equation}
\label{eqn:peevertxeqiv}
p^1= p^2 +x_\gamma
\end{equation}
for some $\gamma\in\Ztre$.
Since we assume that ${\cal P}$ is non-degenerate with respect to 
${\cal L}(g^1,g^2,g^3)$, \eqnn{eqn:peevertxeqiv} implies that
\begin{equation}
\label{eqn:nondegpee}
p^1= p^2.
\end{equation}
Thus any $[v]\in {\cal V}/{\cal L}(g^1,g^2,g^3)$ can be written as
$p+x_\alpha$ for a unique $p\in{\cal P}$: $[v]=[p]$.
Thus there is a natural injection ${\cal V}/{\cal L}(g^1,g^2,g^3)\subset{\cal P}$.
Since $[p]\in{\cal V}$ for all $p\in{\cal P}$, this is an isomorphism.

\begin{figure}
\centerline{(a)\; \includegraphics[width=1.2in,angle=-0]{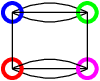}
\qquad      (b)\; \includegraphics[width=1.2in,angle=-0]{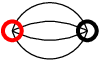}
\qquad      (c)\; \includegraphics[width=1.0in,angle=-0]{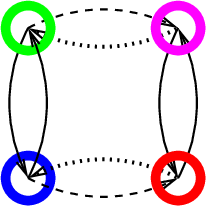}}
\caption{The fundamental finite graphs of ice Ih (a), ice Ic (b), and ice II (c).}
\label{fig:iceonestopol}
\end{figure}

For edges $e\in{\cal E}$, let us use the notation $e_i$ for the vertices of the edge:
$e=(e_1,e_2)$.
We say that $e$ and $\hat e$ are equivalent with respect to the 
group ${\cal L}(g^1,g^2,g^3)$ if, for some $\beta\in\Ztre$,
\begin{equation}
\label{eqn:grapheqiv}
e_i=\hat e_i +x_\beta \Forall i=1,2.
\end{equation}
We define ${\cal E}/{\cal L}(g^1,g^2,g^3)$ to be the set of equivalence classes
with respect to the equivalence relation \eqnn{eqn:grapheqiv}.
Note that $[e] =[\hat e]$ implies that $[e_i] =[\hat e_i]$ for $i=1,2$.
Therefore, for each ``edge'' $[e]\in {\cal E}/{\cal L}(g^1,g^2,g^3)$, we can identify
two vertices $[e_i]$ for $i=1,2$.
Thus there is a natural injection 
\begin{equation}
\label{eqn:equivsubset}
{\cal E}/{\cal L}(g^1,g^2,g^3)\subset
{\cal V}/{\cal L}(g^1,g^2,g^3)\times {\cal V}/{\cal L}(g^1,g^2,g^3)
\subset {\cal P}\times {\cal P}.
\end{equation}
Thus we can think of ${\cal E}/{\cal L}(g^1,g^2,g^3)$ as providing edges between
vertices ${\cal V}/{\cal L}(g^1,g^2,g^3)$, and this defines the graph
${\cal C}/{\cal L}(g^1,g^2,g^3)$, called the \defdex{fundamental finite graph}
of the crystal \cite{ref:K4crystalattice}.
\figgar{fig:iceonestopol}
shows how this graph can distinguish between different crystal structures of ice.

\subsection{Local graph structure}
\label{sec:locrafstruc}

\begin{wrapfigure}{R}{3.2in}
\vspace{-10pt}
\centerline{\includegraphics[width=2.5in]{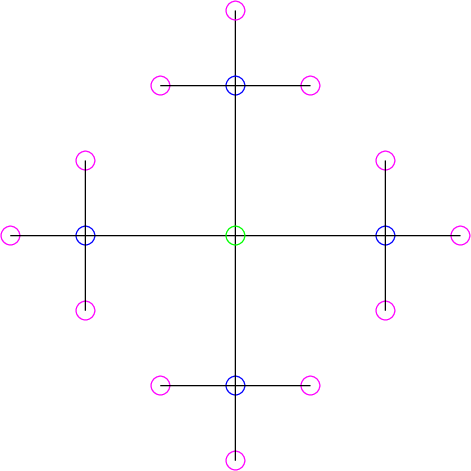}}
\caption{The local graph of hydrogen bonds in both forms of ice I.}
\vspace{-10pt}
\label{fig:icelocalogy}
\end{wrapfigure}

In some cases, the local graph of hydrogen bonds can be instructive.
For crystals, this structure is repeated, so we can look at a single instance.
In \figgar{fig:icelocalogy}, we show the water molecules (vertices indicated by
circles) surrounding the central green water, together with the hydrogen bonds (edges).
The blue waters are the nearest neighbors, and magenta waters are next-nearest neighbors.

Up to this point, both forms of ice I have identical graph topology.
However, as indicated in \figgar{tabl:radisticeo}, there is a difference beyond this, since
the radial distributions functions differ from this point onwards.
In \figgar{fig:onehclocalogy}, the black circles indicate next-next-nearest neighbors.
Notice that four of them are marked in red because they are involved in a cycle.

%% file: iceoneall.tex
\subsection{Ice I structures} 
\label{chap:iceonestruk}

The most common phase of ice on earth is Ih, the hexagonal form of ice
\cite{ref:quantumiceonehstruc,ref:iceIhXIprotonordering}.
However, another form (Ic) also occurs in the atmosphere \cite{ref:iceoneseeatmosfear}.
Both of these have what we call exact tetrahedral structure, as depicted
in \figgar{fig:onecube}.
In this structure, one water molecule is in the center of a square cube, and it
is hydrogen bonded to four water molecules at four corners of the cube, as depicted
in \figgar{fig:onecube}.

It is surprising that two different crystalline structures can be formed from this
basic unit.
One objective here is to examine mathematical tools that can describe how they
differ.
In addition, other forms of ice form nearly tetrahedral structures that have
four bonds, although not with the same lengths and angles as depicted
in \figgar{fig:onecube}.

\section{Ice Ih}
\label{sec:iceonehaa}

The crystal structure of ice Ih is identical to the hexagonal diamond structure of 
carbon, known as Lonsdaleite \cite{ref:hexdiamondshock}.
This structure is also known as wurtzite \cite{ref:coopolarisationiceIh}.
The structure can be visualized in several ways.
It can be viewed as parallel sheets made of a hexagonal network of hydrogen bonds,
as shown in \figgar{fig:iceonehlayer}.
This figure provides the top view of a single sheet in the crystal and depicts
three of the hydrogen bonds formed by each water molecule, each of which is 
represented by a vertex in \figgar{fig:iceonehlayer}(a).
There are alternating colors because the water molecules make alternating hydrogen
bonds with either the sheet above or the sheet below.
In our rendering, the red waters make bonds with waters directly above them, and
the magenta waters make bonds with waters directly below them.
Obscured in \figgar{fig:iceonehlayer}(a) are the vertical undulations in the sheets.
This is made clear when we realize that the hydrogen bonds around each vertex must
form exact tetrahedral bonds with their neighbors, as depicted in \figgar{fig:onecube}.
Thus the red waters, which make bonds with waters directly above them, have all of
the hydrogen bonds in \figgar{fig:iceonehlayer}(a) projecting down.
Thus the red waters are at the top of the undulations in the sheets.
The reverse is true for the magenta waters.

\begin{figure}
\centerline{(a)\includegraphics[width=2.3in]{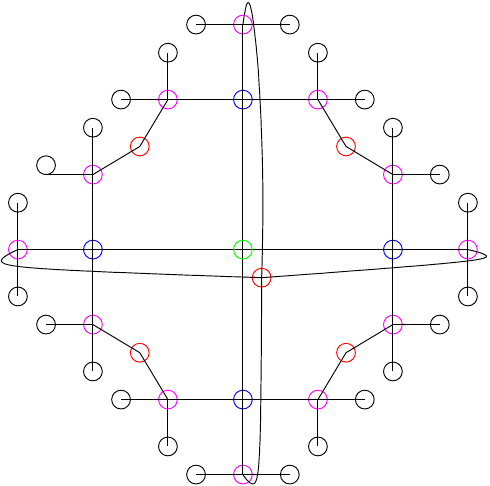}
            (b)\includegraphics[width=2.5in]{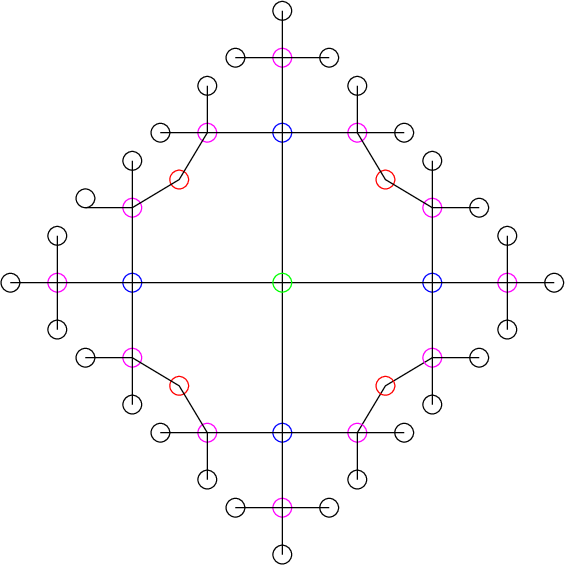}}
\caption{The local graphs of hydrogen bonds in different forms of ice I.
(a) ice Ih, (b) ice Ic.}
\vspace{-10pt}
\label{fig:onehclocalogy}
\end{figure}

We can complete the picture by considering neighboring sheets and their interconnecting
hydrogen bonds, as shown in \figgar{fig:iceonehlayer}(b).
We have drawn the neighboring sheets with blue and green waters to make them distinct
from the red and magenta ones.
We have depicted in \figgar{fig:iceonehlayer}(b) the edge view of a vertical slice 
through the crystal sheets.
Now we see the undulations of the sheets that were obscured 
in \figgar{fig:iceonehlayer}(a).
We can also see that we can view the blue-green sheets as a reflection of the
red-magenta sheets.
The newly revealed hydrogen bonds are depicted in \figgar{fig:iceonehlayer}(b) via
darker lines; the lighter lines indicate hydrogen bonds already seen in 
\figgar{fig:iceonehlayer}(a).

In \figgar{fig:iceonehlayer}(c), the fundamental finite graph 
(\sexshun{sec:quotiegraph}) of the connections \cite{sunada2012topological} is depicted.
This reflects the fact that each water is connected to three waters within its
horizontal sheet and one water in a neighboring sheet.
For example, there are three different ways for a magenta water to be hydrogen
bonded to a red water, and one way for it to be bonded to a green water.

The amplitude of the wiggles in \figgar{fig:iceonehlayer}(b) can be obtained
by reference to \figgar{fig:onecube}.
The plane of the green points passes through the points $(0,2,2),(2,0,2),(2,2,0)$, and
the center of the triangle generated by these points is $(4/3)(1,1,1)$.
Thus the distance from the blue water and the green plane is $\sqrt{3}/3=1/\sqrt{3}$.
Thus the wiggles above and below a mean plane are $\pm c$ where 
$c=1/2\sqrt{3}=(1/6)\sqrt{3}\approx 0.288675$.

\begin{figure}
\centerline{(a)\includegraphics[width=2.0in,angle=-0]{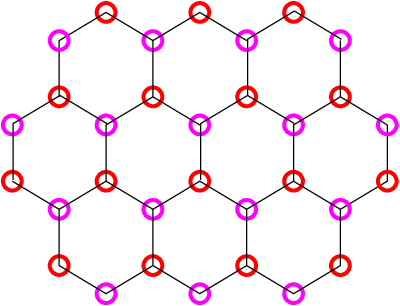}
  \qquad    (b)\; \includegraphics[width=1.0in,angle=-0]{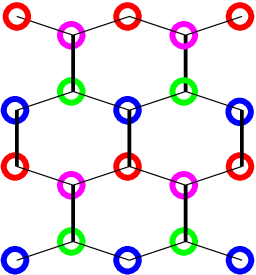}
  \qquad    (c)\; \includegraphics[width=1.2in,angle=-0]{figs/iceonetopolog.eps}}
\caption{The hexagonal structure of ice Ih.
(a) Top view of a single layer.
The red water oxygens are located above the plane of the figure. In addition to the three
hydrogen bonds depicted by lines, they are also hydrogen bonded to waters in another layer
above the plane. The magenta water oxygens are located below 
the plane of the figure and are hydrogen bonded to waters in a layer below the plane.
(b) Side view of the hexagonal structure of ice Ih.
Hydrogen bond linkages between the red-magenta and the blue-green layers are shown 
using a thicker line.
(c) Fundamental finite graph \cite{ref:K4crystalattice,sunada2012topological} of 
the hydrogen bond linkages.
}
\label{fig:iceonehlayer}
\end{figure}

The basic symmetry of Ih ice is known as monoclinic.
The top view of the crystal unit domain is a simple parallelogram, as shown in
\figgar{fig:symmetriceoneh}(a).
In \figgar{fig:symmetriceoneh}(b), data is provided to determine the dimensions
of the crystal unit.
The parameter $a=\sqrt{2}$, since the distance between two next-nearest neighboring 
water oxygens, e.g., two greens in \figgar{fig:symmetriceoneh}(b), is $2\sqrt{2}$.
The blue waters fall in a plane parallel to the plane of \figgar{fig:symmetriceoneh},
so the three dimensional distance is the same as the two dimensional distance.
The parameter $b$ is the side length of a 60 degree right triangle, where the other
side length is $a=\sqrt{2}$.
Thus $b=\sqrt{6}$.

\subsection{Ice Ih water positions}

\begin{figure}
\centerline{(a)\includegraphics[width=3.0in,angle=-0]{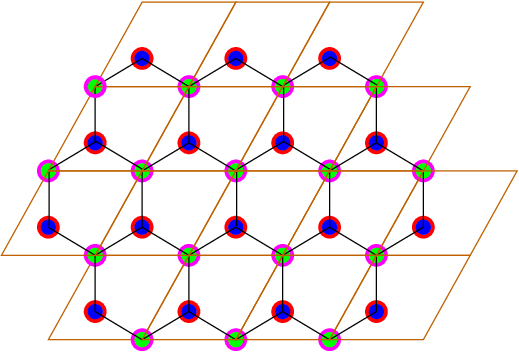}
           (b)\;\includegraphics[width=1.8in,angle=-0]{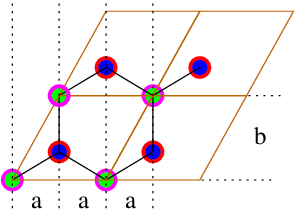}
           (c)\;\includegraphics[width=1.2in,angle=-0]{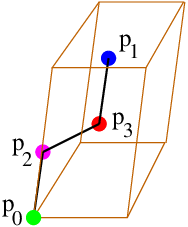}}
\caption{The monoclinic symmetry of ice Ih. (a) The repeating crystal unit in
the $(x,y)$-plane, generated by $g_1$ and $g_2$.
(b) Notation to determine the distance $a$; note that $b=\sqrt{3}\,a$.
(c) Three-dimensional unit cell showing the four water molecules and the edges
connecting them that fall completely within the unit cell.}
\label{fig:symmetriceoneh}
\end{figure}

We recall the parameter $c=(1/6)\sqrt{3}$ for the amplitude of the `wiggle' below 
and above the plane of the figure in \figgar{fig:symmetriceoneh}.
Thus the points in the unit cell in the plane of the figure in 
\figgar{fig:symmetriceoneh} are ($p_0$ is green, $\hat p_1$ is blue)
\begin{equation}
\label{eqn:firsticeIhlocs}
\begin{split}
p_0=& (0,0,0), \\
\hat p_1=&(a,b/3,-2c)=(\sqrt{2},\sqrt{2}/\sqrt{3},-(1/3)\sqrt{3})\approx(1.41,0.82,-0.58).\\
\end{split}
\end{equation}
We can check the values of $a,b,c$ by using the fact that the distance between
oxygen centers is $\sqrt{3}$, that is, $\norm{\hat p_1-p_0}=\sqrt{3}$.
Thus
$$\norm{\hat p_1-p_0}^2=a^2+(b/3)^2+(2c)^2=2+\frac{2}{3} +\frac{1}{3}=3.$$ 
We will see that $\hat p_1$ is not in the crystal unit domain, so will have to make
a small modification.

The remaining points in the unit cell ($p_2$ is magenta, $p_3$ is red) are 
\begin{equation}
\label{eqn:iceIhlocations}
\begin{split}
p_2=&(0,0,\sqrt{3})\approx (0,0,1.73),\\
p_3=&(\sqrt{2},\sqrt{2}/\sqrt{3},(4/3)\sqrt{3})\approx(1.41,0.82,2.31),\\
\end{split}
\end{equation}
which can be determined as follows.
Green and magenta waters ($p_0$ and $p_2$) are hydrogen bonded, and thus they are at a
distance $\sqrt{3}$, and moreover their $x$ and $y$ coordinates are the same.
Thus $p_2-p_0=(0,0,\sqrt{3})$.
The red waters are above the magenta waters by an amount $2c=1/\sqrt{3}=\sqrt{3}/3$.

Note that 
\begin{equation}
\label{eqn:nerernbr}
\norm{p_3-\hat p_1}^2=3\left(\frac{3}{2}+\frac{1}{6}\right)^2=
3\left(\frac{5}{3}\right)^2=\frac{25}{3},
\end{equation}
so that $\norm{p_3-\hat p_1}=5/\sqrt{3}\approx 2.88675$ (see \figgar{tabl:radisticeo}), 
slightly larger than the next-nearest neighbor distance of $2\sqrt{2}\approx 2.828427$.

\begin{figure}
\centerline{(a)\;\includegraphics[width=1.9in,angle=-0]{figs/newsymuniticeoneh.eps}
\qquad      (b)\includegraphics[width=2.5in,angle=-0]{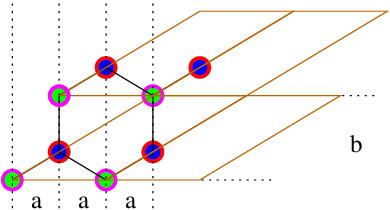}}
\caption{Different crystal units for ice Ih.}
\label{fig:comparriceoneh}
\end{figure}

The generators of the lattice for the infinite crystal as in \defnn{def:crystal} 
for the ice Ih crystal are therefore
\begin{equation}
\label{eqn:iceIhgeneratrs}
\begin{split}
g^1 &=(2a,0,0)=(2\sqrt{2},0,0)\approx (2.83,0,0),\\
g^2 &=(a,b,0)=(\sqrt{2},\sqrt{6},0)\approx (1.41,2.45,0),\\
g^3 &=(0,0,(8/3)\sqrt{3})\approx (0,0,4.62),
\end{split}
\end{equation}
where the $z$-axis translation is easily determined from \figgar{fig:iceonehlayer}(b),
as follows.
The crystal unit repeat distance in the $z$-direction consists of two components:
one is the distance between oxygen centers $\sqrt{3}$ and the other is the oscillation 
in the layer, which has magnitude $c=1/2\sqrt{3}$.
Both of these elements get repeated twice in the crystal unit.
Thus $g^3_3=2\sqrt{3}+2/\sqrt{3}=8/\sqrt{3}$.
An internal check is available regarding the volume of the unit cell,
\begin{equation}
\label{eqn:checkonehvol}
2a\cdot b \cdot g^3_3=(2\sqrt{2})(\sqrt{6})(8/\sqrt{3})=32.
\end{equation}
This means that the water density is one water per 8 cubic units.
We will see that this matches exactly the density of ice Ic.

The point 
\begin{equation}
\label{eqn:remainingpoint}
p_1=\hat p_1 + g^3 = (\sqrt{2},\sqrt{2}/\sqrt{3},(15/6)\sqrt{3})
\end{equation}
is in the crystal unit domain.
Thus the four points, $p_0$, $p_1$, $p_2$, $p_3$, form the generators for the Ih crystal lattice.

\subsection{Orthorhombic ice Ih}

There is another description of ice Ih as a crystal lattice, with a rectangular box
as the unit cell and eight water molecules in the fundamental domain.
The ice XI crystal (see section \ref{sec:iceleven}) is a particular case of ice Ih
in which all of the dipole directions are fixed.
The customary axes for crystals are labeled $a,b,c$.
For ice XI at 5 degrees Kelvin, the lengths of the fundamental domain sides are
$a\approx 4.465$\AA, $b\approx 7.859$\AA, and $c\approx 7.292$\AA\ \cite{ref:quantumiceonehstruc}.
The structure is depicted in \cite[Figure 4]{ref:quantumiceonehstruc}.

\begin{figure}
\centerline{(A)\includegraphics[width=1.8in]{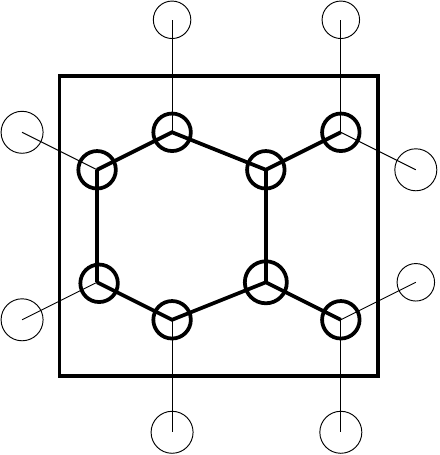}
\qquad\qquad(B)\includegraphics[width=1.8in]{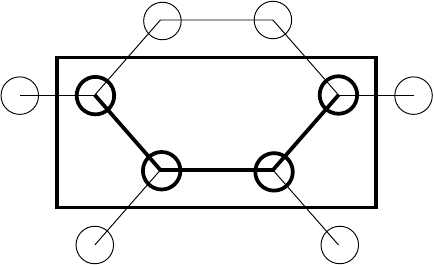}}
\caption{The orthorombic description of ice Ih and ice XI.
(A) The $b-c$ plane of ice XI.
(B) The $a-b$ plane of ice XI.}
\label{fig:iceleven}
\end{figure}

In Figure \ref{fig:iceleven} we give a caricature of \cite[Figure 4]{ref:quantumiceonehstruc}.
In panel (A) of Figure \ref{fig:iceleven} we give the view of the $b-c$ plane, looking
along the $a$ axis.
In panel (B) of Figure \ref{fig:iceleven} we give the view of the $a-b$ plane, looking
along the $c$ axis.
By symmetry, the boundary planes of the fundamental domain always bisect
the hydrogen bonds that go outside of the fundamental domain.
Thus all hydrogens associated with the eight oxygens in the fundamental domain
are themselves contained in the fundamental domain.

\subsection{Ice Ih sheets}

If we consider the two dimensional crystal lattice generated by the vectors $g^1$ and $g^2$
and the points $p_0$ and $\hat p_1$, we see that it corresponds to the sheet depicted in
\figgar{fig:symmetriceoneh}(a).

\subsection{Ice Ih graph edges}

\begin{wrapfigure}{R}{1.7in}
\centerline{\includegraphics[width=1.5in]{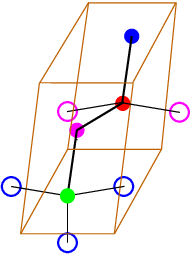}}
\caption{The edges ${\cal S}$ that generate the ice Ih crystal lattice via \eqnn{eqn:crystaledge}.}
\label{fig:edgesiceoneh}
\end{wrapfigure}

The edges of the ice Ih crystal lattice can be deduced from the vertices since they represent the
nearest-neighbor connections.
However, this is not true of all ice crystal lattices.
So it is useful to describe a set of edges ${\cal S}$ that generate the ice Ih crystal
lattice via \eqnn{eqn:crystaledge}.
First of all, let us count how many edges are required.
From \figgar{fig:iceonehlayer}, we see that for each face of the unit cell that is depicted as
a line in \figgar{fig:iceonehlayer}(a), there are two edges crossing, one for each green-blue pair
and one for each magenta-red pair.
This accounts for 8 edges per unit cell.
In addition, in the vertical direction normal to the plane in \figgar{fig:iceonehlayer}(a), there is
only one edge per face, one for each blue-red pair.
This adds two more edges per unit cell, for a running total of 10.
Interior to the unit cell, there are three more edges as shown in \figgar{fig:iceonehlayer}(c).
This brings the total to 13 edges per unit cell.
Of these edges, all of the ten that cross a face will be counted twice.
Thus the number of edges per unit cell is 8.
Therefore we expect to need at least 8 edges in ${\cal S}$.

In \figgar{fig:edgesiceoneh}, we depict 8 edges that can be used to generate all of the
edges of the ice Ih crystal lattice via \eqnn{eqn:crystaledge}.
Since the crystal unit domain has volume 32, according to \eqnn{eqn:checkonehvol},
this means there is one hydrogen bond per 4 cubic units.

A generator set ${\cal S}$ for the edges for the infinite crystal as in \defnn{def:crystal}
for the ice Ih crystal consists of
\begin{equation}
\label{eqn:iceIhgenedges}
\begin{split}
&(p_0,p_1-g^1-g^3),\quad
(p_0,p_1-g^2-g^3),\quad
(p_0,p_1-g^3),\quad
(p_0,p_2),\\
&(p_2,p_3),\quad
(p_3,p_1),\quad
(p_3,p_2+g^1+g^2),\quad
(p_3,p_2-g^1+g^2).
\end{split}
\end{equation}

\section{Ice Ic}
\label{sec:iceonesee}

\begin{figure}
\centerline{(a)\includegraphics[width=1.6in]{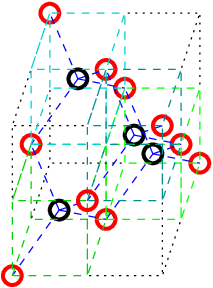}
\qquad   (b)\includegraphics[width=1.3in]{figs/icediamondtop.eps}}
\caption{Crystal structure of ice Ic. (a) Cubic lattice unit of ice Ic.
(b) Fundamental finite graph \cite{ref:K4crystalattice} of the hydrogen bond 
linkages.}
\label{fig:cubicunit}
\end{figure}

The location of water molecules in the ice Ic lattice is the same as the
diamond (carbon) lattice.
The placement of waters in a cube is indicated in \figgar{fig:onecube}.
In \figgar{fig:cubicunit}, we depict water molecules in a cube of side four.
This has four sub-cubes of size two each, each containing a molecule at its center,
as well as four void cubes.
The symmetry of the diamond (ice Ic) crystal is often called face-centered cubic,
since there are atoms at each corner of the cube of side four and at 
the center of each face.

The location of the waters ${\cal P}$ in the unit cube that can be used to generate the
infinite crystal as in \defnn{def:crystal} are
\begin{equation}
\label{eqn:iceIclocations}
\begin{split}
p_0=000,\;
p_1=111,\;
p_2=& 220,\;
p_3=022,\;
p_4=202,\; \\
p_5=& 331,\;
p_6=133,\;
p_7=313,
\end{split}
\end{equation}
where we use the abbreviation $xyz$ to stand for $(x,y,z)$ for vectors in $\Rtre$,
in keeping with the notation in \figgar{fig:onecube}.
Notice that the last three vectors, which correspond to the subcube centers
in \figgar{fig:cubicunit} other than 111, can be written as the vector sum of 111
and the third, fourth and fifth vectors, the positions of the waters at the 
corners of the subcube in \figgar{fig:onecube}.
That is,
\begin{equation}
\label{eqn:iceIcrelations}
p_i=p_{i-3}+p_1,\; i=5,6,7.
\end{equation}
The generators of the lattice for the infinite crystal as in \defnn{def:crystal} 
for the ice Ic crystal are
\begin{equation}
\label{eqn:iceIcgeneratrs}
g^1=400,\quad
g^2=040,\quad
g^3=004.
\end{equation}
The density of of ice Ic is eight molecules in 64 units cubed, or one
molecule per 8 cubic units.

\begin{wrapfigure}{R}{3.4in}
\centerline{\includegraphics[width=3.2in]{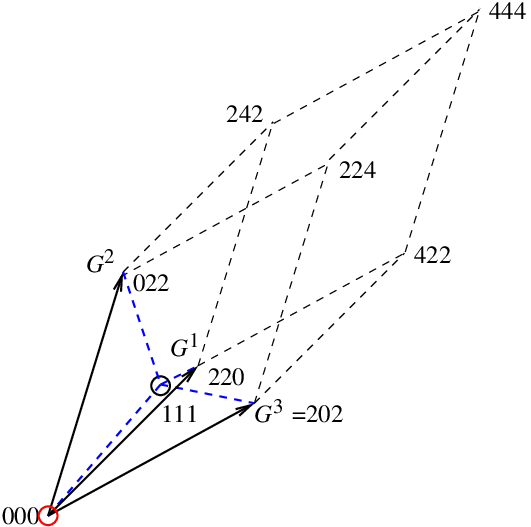}}
\caption{Smaller unit domain for crystal structure of ice Ic.
The volume of the smaller unit is 16 cubic units.}
\vspace{-10pt}
\label{fig:onecunit}
\end{wrapfigure}

The relation \eqnn{eqn:iceIcrelations} suggests that the diamond (ice Ic) crystal can be 
generated by a smaller number of generators \cite{ref:K4crystalattice,sunada2012topological}.
So we consider crystal generators
\begin{equation}
\label{eqn:tugeniceIclocs}
\begin{split}
P_0=p_0=000,\quad
P_1=p_1=111,
\end{split}
\end{equation}
together with lattice generators
\begin{equation}
\label{eqn:togeniceIclatis}
G^1=p_2= 220,\quad
G^2=p_3=022,\quad
G^3=p_4=202.
\end{equation}
Then \eqnn{eqn:iceIcrelations} implies that
\begin{equation}
\label{eqn:newiceIcrels}
p_{i+3}=G^i+p_1,\; i=2,3,4.
\end{equation}
Of course we also have
\begin{equation}
\label{eqn:trivialcrels}
p_{i+1}=G^i+p_0,\; i=1,2,3.
\end{equation}
Similarly, we have
\begin{equation}
\label{eqn:latgenIcrels}
g^{1}=G^1-G^2+G^3,\quad
g^{2}=G^1+G^2-G^3,\quad
g^{3}=-G^1+G^2+G^3.
\end{equation}
Thus we have shown that the set of crystal vertices generated from
\eqnn{eqn:togeniceIclatis} and \eqnn{eqn:tugeniceIclocs} contain all of
the crystal vertices generated from
\eqnn{eqn:iceIcgeneratrs} and \eqnn{eqn:iceIclocations}.
The crystal unit domain is depicted in \figgar{fig:onecunit}.

The fact that the crystal with a smaller number of generators does not generate
more points can be proved by a density argument.
The volume of the fundamental domain of the diamond crystal can be computed as follows.
We use the change of variables \eqnn{eqn:latgenIcrels}, which we can write as
\begin{equation}
\label{eqn:rewlatgenIcrels}
\begin{pmatrix} g^{1} \\ g^{2} \\ g^{3}\end{pmatrix}=
\begin{pmatrix} 1 & -1 & 1 \\ 1 & 1 & -1 \\ -1 & 1 & 1 \end{pmatrix} 
\begin{pmatrix} G^{1} \\ G^{2} \\ G^{3}\end{pmatrix} .
\end{equation}
Thus the fundamental domain for the lattice ${\cal L}(G^{1},G^{2},G^{3})$
has volume $16=64/4$, since
\begin{equation}
\label{eqn:fundomanvol}
\hbox{det} \begin{pmatrix} 1 & -1 & 1 \\ 1 & 1 & -1 \\ -1 & 1 & 1 \end{pmatrix} = 4.
\end{equation}
Thus the density of molecules in the crystal generated from
\eqnn{eqn:togeniceIclatis} and \eqnn{eqn:tugeniceIclocs} is again one molecule
per eight cubic units.
Therefore the two crystals are the same.

This confirms that the fundamental finite graph of the diamond (ice Ic) crystal is 
as shown in \figgar{fig:cubicunit}(b) \cite{ref:K4crystalattice,sunada2012topological}.

\subsection{Ice Ic sheets}

We now prove that ice Ic can also be represented by sheets as depicted in
\figgar{fig:symmetriceoneh}(a).
Comparing \figgar{fig:onecunit} and \figgar{fig:symmetriceoneh}(c), aligning $p_2$ in the
latter with $p_1$ in the former, we see that the sheet in ice Ic is orthogonal to the
direction of $p_1$.
Thus we consider the set of points
$$
\bigset{P_j+\sum_{i=1}^3 \alpha_i G^i}{j=0,1; \sum_{i=1}^3 \alpha_i=0,\;\alpha\in\Ztre},
$$
where we recall that $P_0=(0,0,0)$ and $P_1=(1,1,1)$.

If we consider the two dimensional crystal lattice generated by the vectors $g^1$ and $g^2$
and the points $p_0$ and $\hat p_1$, we see that it corresponds to the sheet depicted in
\figgar{fig:symmetriceoneh}(a).

\begin{figure}
\centerline{(a) \includegraphics[width=2.0in,angle=-0]{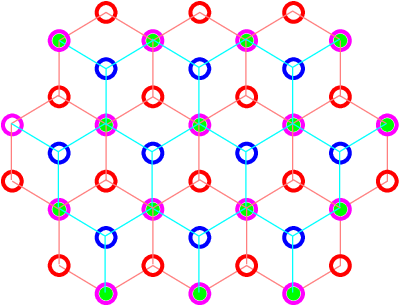}
\qquad  (b) \includegraphics[width=2.5in,angle=-0]{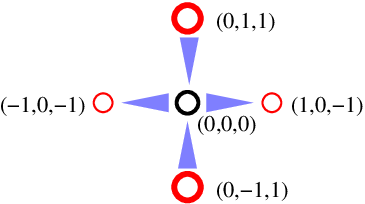}}
\caption{The hexagonal view of the ice Ic crystal structure.
(a) Overlay of two layers. The bottom blue-green layer is shifted so that only
the green waters lie directly below (and are hydrogen bonded to) the magenta
waters.
The blue waters are not hydrogen bonded to any of the waters in the red-magenta
layer shown, but rather are hydrogen bonded to waters in a layer (not shown)
below the blue-green layer.
(b) The basic tetrahedral unit. The blue triangles indicate hydrogen bonds, and
the direction of the triangle indicates the direction in or out of the plane.
}
\label{fig:tulayriceone}
\end{figure}

\subsection{Ice Ic graph edges}

If the number of hydrogen bonds per unit volume is the same for Ic and Ih ice, then we expect
there to be four hydrogen bonds in the crystal unit domain for Ic.
There are four such edges inside each unit domain (the blue dashed lines in \figgar{fig:onecunit}).
The corresponding generator set ${\cal S}$ for the edges for the infinite ice Ic crystal 
via \defnn{def:crystal} consists of
\begin{equation}
\label{eqn:iceIcgenedgeset}
(p_0,p_1),\quad
(p_1,p_2),\quad
(p_1,p_3),\quad
(p_1,p_4).
\end{equation}

\subsection{Second view of the Ic crystal structure}
\label{sec:iceoneshift}

The ice Ic (diamond) crystal structure is closely related to the Ih structure of ice.
It involves the same hexagonal sheets as indicated in \figgar{fig:iceonehlayer}.
In this case, there are three such sets of parallel layers, running in transverse
directions.
So to be precise, we must pick one normal direction for definiteness, which we
will call the $z$ axis.
In the case of ice Ih, there is only one normal direction to the parallel
sheets determined by the direction of the regular hexagonal holes.

For ice Ic, instead of hydrogen bonding to reflected layers above and below, they are
hydrogen bonded to layers above and below that are shifted, as is shown in
\figgar{fig:tulayriceone}(a) for two layers.
We imagine that the blue-green layer is below the red-magenta layer.
Thus we see the green waters below the magenta waters.
The red waters in the red-magenta layer are also hydrogen bonded to blue waters
in an unseen blue-green layer above.
Thus the two blue-green layers are shifted as well and do not lie on top of each
other from this perspective.

There is also another description of the ice Ic (diamond) crystal, involving a smaller
set of generators and lattice unit domain.
We pick a  set of coordinates $(\hat x,\hat y,z)$ rotated by 45 degrees in the 
plane perpendicular to the $z$-axis.
In these coordinates, the location of the waters ${\cal P}$ in the unit cube that 
can be used to generate the infinite crystal as in \defnn{def:crystal} are
\begin{equation}
\label{eqn:iceIslocs}
\hat p_0=(0,0,0),\quad
\hat p_1=(\sqrt{2},0,1),\quad
\hat p_2=(\sqrt{2},\sqrt{2},2),\quad
\hat p_3=(0,\sqrt{2},3).
\end{equation}
The generators of the lattice for the infinite crystal as in \defnn{def:crystal}
for the Ic crystal in these coordinates are
\begin{equation}
\label{eqn:iceIsgeners}
\hat g^1=(2\sqrt{2},0,0),\quad
\hat g^2=(0,2\sqrt{2},0),\quad
\hat g^3=(0,0,4).
\end{equation}

\begin{figure}
\centerline{(a) \includegraphics[width=2.0in,angle=-0]{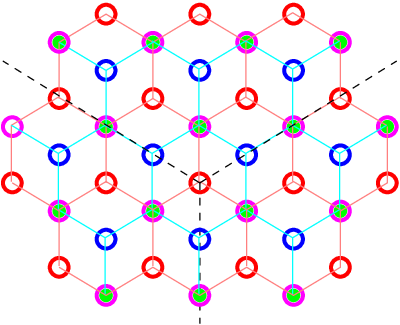}
  \qquad    (b)\; \includegraphics[width=1.0in,angle=-0]{figs/iceonehsidvu.eps}}
\caption{Coordinate axes for the ice Ic crystal structure.
(a) Overlay of two layers. The bottom blue-green layer is shifted so that only
the green waters lie directly below (and are hydrogen bonded to) the magenta
waters.
The blue waters are not hydrogen bonded to any of the waters in the red-magenta
layer shown, but rather are hydrogen bonded to waters in a layer (not shown)
below the blue-green layer.
(b) Side view.}
\label{fig:onecaxes}
\end{figure}

The relation between the coordinates in \figgar{fig:onecube} and the points in
\eqnn{eqn:iceIsgeners} and \eqnn{eqn:iceIslocs} is given by
\begin{equation}
\label{eqn:changcoord}
x=\frac{\hat x-\hat y}{\sqrt{2}},\quad
y=\frac{\hat x+\hat y}{\sqrt{2}}.
\end{equation}
Thus the $(x,y,z)$ coordinates $q_i$ of $\hat p_i$ compare with the points $p_i$
of \eqnn{eqn:iceIclocations} as follows:
\begin{equation}
\label{eqn:unhatpees}
\begin{split}
q_0=&(0,0,0)=p_0,\quad
q_1=(1,1,1)=p_1,\quad
q_2=(0,2,2)=p_3,\\
q_3=&(-1,1,3)=(3,1,3)-(4,0,0)=p_7-g^1.
\end{split}
\end{equation}
Similarly, the $(x,y,z)$ coordinates $h^i$ of the lattice 
vectors $\hat g^i$ are
\begin{equation}
\label{eqn:unhatgees}
h^1=(2,2,0)=\half(g^1+g^2),\quad
h^2=(-2,2,0)=\half(-g^1+g^2),\quad
h^3=g^3.
\end{equation}
Inverting these relations, we find
\begin{equation}
\label{eqn:nverthatgees}
g^1=h^1-h^2,\quad
g^2=h^1+h^2,\quad
g^3=h^3.
\end{equation}
Thus any point of the form in the ice Ic crystal can be written as
\begin{equation}
\label{eqn:genformcrys}
p_j+\sum_{i=1}^3 m_i g^i=p_j+(m_1+m_2)h^1+(m_2-m_1)h^1+m_3 h^3.
\end{equation}
To see that this is in the crystal described by 
\eqnn{eqn:iceIslocs} and \eqnn{eqn:iceIsgeners}, we have to relate the rest
of the $p_i$'s to the $q_i$'s.
Thus we collect all of the relationships here, some of which have previously
been derived:
\begin{equation}
\label{eqn:mopeerel}
\begin{split}
p_0=&q_0\\
p_1=&q_1\\
p_2=&(2,2,0)=h^1=q_0+h^1 \\
p_3=&q_2\\
p_4=&(2,0,2)=q_2-h^2 \\
p_5=&(3,3,1)=q_1+h^1 \\
p_6=&(1,3,3)=q_3+h_1 \\
p_7=&q_3+g^1 =q_3+h^1 -h^2 . \\
\end{split}
\end{equation}
This proves that 
\begin{equation}
\label{eqn:cryscontain}
\begin{split}
{\cal V}(p_0,\dots,p_7;g^1,g^2,g^3)\subset
{\cal V}(q_0,\dots,q_3;h^1,h^2,h^3)=
\widehat{\cal V}(\hat p_0,\dots,\hat p_3;\hat g^1,\hat g^2,\hat g^3).
\end{split}
\end{equation}
The fact that 
\begin{equation}
\label{eqn:crysequalin}
\begin{split}
{\cal V}(p_0,\dots,p_7;g^1,g^2,g^3)=
{\cal V}(q_0,\dots,q_3;h^1,h^2,h^3)=
\widehat{\cal V}(\hat p_0,\dots,\hat p_3;\hat g^1,\hat g^2,\hat g^3)
\end{split}
\end{equation}
is a simple consequence of a density argument.
The former crystal has eight points per cube of size $4^3=64$ units-cubed, whereas 
the latter has four points per box of volume 
$2\sqrt{2}\times2\sqrt{2}\times 4=32$ units-cubed.
Thus they both have the same density of one point per eight units-cubed, and hence
they must be equal.
If some point were in the latter and not in the former, this would happen consistently
(periodically) and this would violate the equality of densities.

\subsection{Alternating Ih/Ic layered structures}
\label{sec:althslayers}

\begin{figure}
\centerline{(a) \includegraphics[width=2.0in,angle=-0]{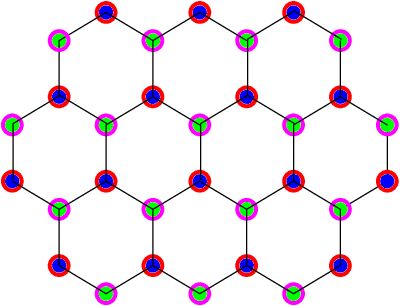}
\qquad      (b) \includegraphics[width=2.0in,angle=-0]{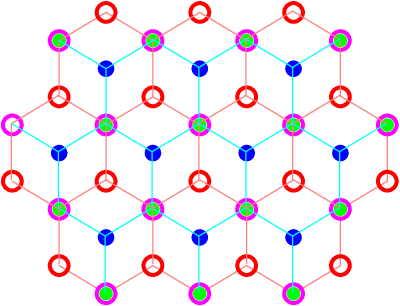}}
\caption{Comparison of the hexagonal views of the ice Ih (a) and Ic (b) crystal structures.
In both structures, the blue waters are not hydrogen bonded to any of the waters 
in the red-magenta layer shown, but rather are hydrogen bonded to waters in a 
layer (not shown) below the blue-green layer.
The magenta and green waters are hydrogen bonded.
}
\label{fig:concononehonec}
\end{figure}

The two ice I structures are compared and contrasted in \figgar{fig:concononehonec}.
Since both the Ih and Ic crystals can be constructed one layer at a time,
it is clear that layered structures can be built with arbitrary alternations
between Ih and Ic layers.
More precisely, when adding an additional layer, a choice can be made to add
a reflection (Ih) or shift (Ic) of the previous layer.
This construction preserves the exact local tetrahedral structure depicted in
\figgar{fig:onecube} or \figgar{fig:tulayriceone}(b).
Thus one set of possible ice I structures can be identified with the set
of doubly infinite sequences of binary characters, .e.g., ....ccccchccchhccchhcchhhhhhhcc.....
Ice structures of this type are called stacking-disordered ice (ice I$_{\rm sd}$)
\cite{ref:stackdisordericeI}.
These structures may contribute to the ambiguity seen in phase diagrams for ice where the
boundary between Ih and Ic is given as a dashed line.

%% file: icetwo.tex
\section{Ice II structure} 
\label{chap:icetustruk}

\begin{figure}
\centerline{\includegraphics[width=2.0in,angle=-90]{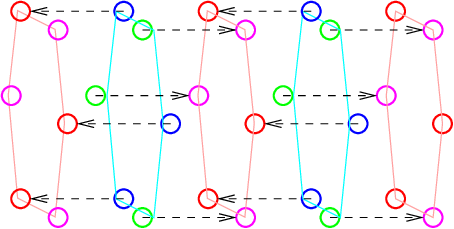}}
\caption{The structure of the hexagonal columns in ice II.
Side view of hydrogen bond structure between the two types of rings in ice II.
Oscillations in the oxygen displacements from the mean are indicated for the blue-green
hexagon; the solid line hexagon depicts the mean position along the axis of the
hexagonal column.}
\label{fig:icetwosidevu}
\end{figure}

The structure of ice II \cite{ref:iceIIxraydata} is different from the ice I 
structures in several ways, but it can also be compared with ice Ih in useful ways.
It is useful to think of ice II as being derived from ice Ih in terms 
of hexagonal columns, each of which is formed of an alternating structure 
of hexagonal rings as depicted in \figgar{fig:icetwosidevu}.
The columns themselves no longer form simple hexagonal attachments, however.
Note that the alternating hexagons are not perfectly aligned in ice II, cf.~Figure 1 
in \cite{ref:iceIIweakenedHbond} or Figure 2a in \cite{ref:D2Oicetwopowder}.

\subsection{Two ring types}

\begin{figure}
\centerline{\includegraphics[width=6.0in,angle=-0]{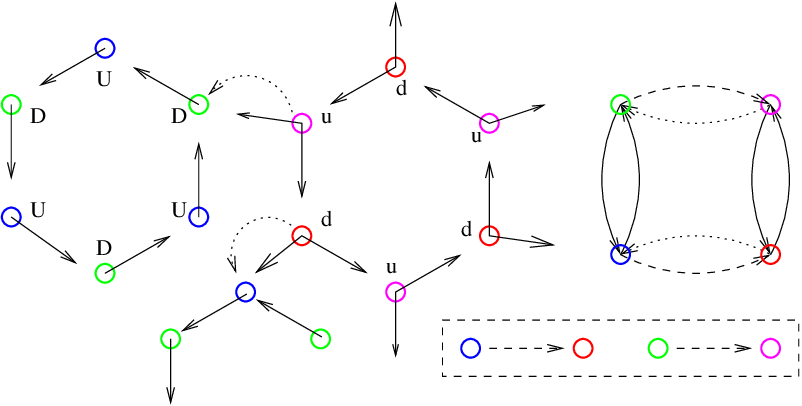}}
\caption{Hydrogen bond structure in the two types of rings in ice II.
The dashed lines in the dashed box indicate the hydrogen bonds made perpendicular to
the plane of the figure, whose directions are indicated by U and D (up and down).
The lower case letters `u' and `d' (up and down) indicate the orientation of 
the hydrogen bond acceptor regions on the oxygens.
The dotted lines indicate hydrogen bonds made between two different hexagonal columns.
The graph to the right depicts the topology of the hydrogen bond connections
\cite{ref:K4crystalattice}.}
\label{fig:icetwo}
\end{figure}

Thus there are two types of rings: the blue-green rings form hydrogen bonds to the
red-magenta rings above and below, much like the connections formed between layers
in ice Ih.
However, the layers in this case are quite different, as the red-magenta rings do
not make hydrogen bonds with the blue-green rings directly above or below them.
Instead, the red-magenta rings form hydrogen bonds with neighboring hexagonal
columns, as depicted in \figgar{fig:icetwo}.

Unlike ice Ih and Ic, ice II is hydrogen-ordered (or proton-ordered).
This means there is exactly one prescribed direction for each of the hydrogen bonds
formed at each oxygen center.
Thus we can depict the hydrogen bonds via an oriented graph, as indicated
in \figgar{fig:icetwo}.
Also, the local structure of ice II is significantly altered from the perfect
tetrahedral structure of ice Ih and Ic.
This is consistent with the fact that water is itself not tetrahedral in structure
\cite{ref:finneywatermolecule}; it is trigonal, but the flexibility of hydrogen
bonds allows the formation of both perfect and approximate tetrahedral structures.

For ice II, the hexagonal columns are somewhat free-standing, connected by hydrogen
bonds to other hexagons, but the connections themselves do not form a hexagonal
structure as in ice Ih.
Each of the hexagonal columns is constructed using alternating layers of hexagons
as depicted in \figgar{fig:icetwosidevu}.
Thus there are green-blue hexagons and red-magenta hexagons which stack on top
of each other.
Oscillations in the water heights around the mean are indicated for the blue-green
hexagon; the solid line hexagon depicts the mean height.
The units used in \cite{ref:iceIIxraydata} are based on the unit of repeat in the 
direction orthogonal to the plane of \figgar{fig:icetwo}, which is the same as the 
direction parallel to the center of the hexagonal columns in \figgar{fig:icetwosidevu}
(i.e., the horizontal axis); one of these units is approximately 6.25 \AA.
The length of the hydrogen bonds shown in \figgar{fig:icetwosidevu} corresponds to 
about 0.57 units in the direction parallel to the center of the hexagonal columns, 
whereas the distances between nonbonded oxygens is only 0.43 units.
In \figgar{fig:icetwo}, which is a caricature of Figure 2 in \cite{ref:iceIIxraydata},
the top view of each of these hexagons is shown, together with the hydrogen bonds 
made between two different hexagonal columns, indicated by a dotted line.

\begin{figure}
\centerline{\includegraphics[width=3.0in,angle=-0]{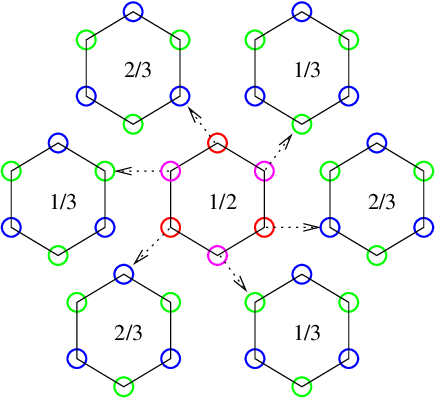}
\qquad      \includegraphics[width=3.0in,angle=-0]{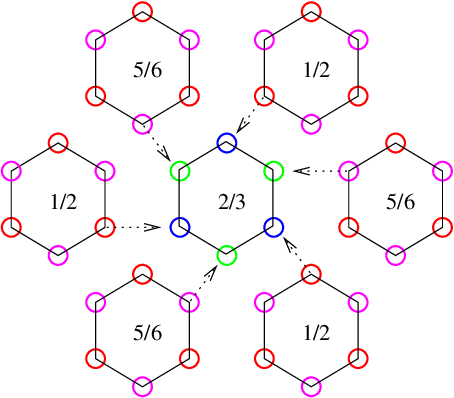}}
\caption{Schematic of one class of hydrogen bonds in ice II
that link different hexagonal columns.
The fractions in each hexagon indicate the mean elevation of the hexagon.}
\label{fig:manytubicetwo}
\end{figure}

The alternating colors mark the alternating directions of the hydrogen bonds (from
the oxygen center towards the two hydrogens in each water molecule).
In the left hexagon, only one of the hydrogen bonds is seen as the other is 
(alternately) going out of the page or into the page; the directions for these 
hydrogen bonds are indicated by capital letters for Up and Down.
The pattern of these hydrogen bonds is indicated by the dashed lines in the 
dashed box: e.g., green waters have hydrogens below the plane of the page 
making a hydrogen bond with magenta oxygens.
The small letters indicating `u' and `d' in the red-magenta hexagons also indicate
the directions of the hydrogen bonds, but in this case in reverse. 
These hydrogen bonds are depicted by arrows that are visible in the plane of the page; 
the larger arrows indicate ones coming up out of the page, and
the smaller arrows indicate ones going down.
Thus the letter `u' does not indicate the direction of this hydrogen bond, but rather
the direction of the donor pair for the hydrogen bond connecting to it.
Thus the magenta water oxygen is the acceptor for the hydrogen bond to the green 
water above it.
Note that the `u' does indicate the position of the acceptor, which is up since the
hydrogen bond emanating from the magenta water is down.
So the capital letters U and D indicate directions of the hydrogen bond donor pairs,
whereas the lower case letters indicate the orientation of the oxygen acceptor region.

\begin{figure}
\centerline{\includegraphics[width=3.0in,angle=-0]{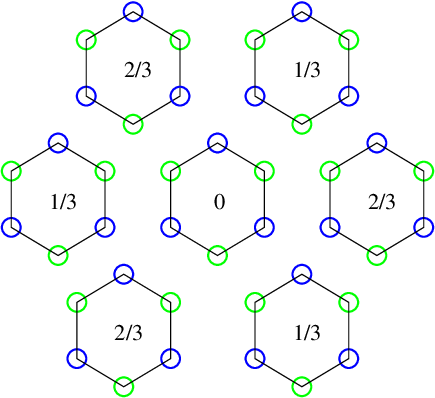}}
\caption{Top view of alternating elevations of nearby blue-green hexagons in ice II.
For every hexagon at elevation $2/3$, there is one below it at elevation $-1/3$.}
\label{fig:altreicetu}
\end{figure}

\subsection{Planarity}

The red-magenta hexagons are very nearly planar, but the green-blue hexagons are
less so.
This allows the out-of-plane hydrogen bonds to be more nearly orthogonal to the
plane of \figgar{fig:icetwo}.
The units used in \cite{ref:iceIIxraydata} are based on the unit of repeat in
the direction orthogonal to the plane of \figgar{fig:icetwo}.
In these units, the oscillation of the red and magenta oxygen centers is 
about $\pm 0.02$ around the mean plane, whereas the green and blue oxygen 
centers are about $\pm 0.05$ from the mean plane.
The greens and reds are below their means, and the magentas and blues 
are above their means.

The dotted lines in \figgar{fig:icetwo} indicate the sideways 
connections between hexagonal tubes.
To clarify the picture, we indicate all the sideways hydrogen bonds emanating 
from one hexagon on the left in \figgar{fig:manytubicetwo}.
To complete the picture, we have indicated all of the acceptors of hydrogen bonds 
for the green-blue waters.
These cannot be easily combined in a planar plot, one key feature of ice II.
It is worth noting that Figure 2b in \cite{ref:D2Oicetwopowder} indicates only one 
direction for the linkages between hexagonal columns, whereas \figgar{fig:manytubicetwo}
shows that they go both up and down (if they did not go both ways, there would be 
significant compressibility of the ice II hexagonal columns).

In \figgar{fig:manytubicetwo}, we have indicated the mean height of the various
hexagons in the units just introduced.
We see that there is an oscillation in mean heights of the connected hexagons.
The alternation in elevation between hexagonal columns is depicted in 
\figgar{fig:altreicetu}, which shows the elevations of nearby blue-green hexagons.
For every hexagon at elevation $2/3$, there is one below it at elevation $-1/3$.
Thus we can view the variation as an oscillation of size $\pm 1/3$.

\section{Ice XI}
\label{sec:iceleven}

The name ice XI appeared in western journals 
in 1986 \cite{ref:koddopediceXIent,ref:Howe1988PhDiceXI}, but had appeared
a year earlier in Japan \cite{suga1985phase}.
Ice XI has the same lattice structure as ice I, and there are two forms,
ice XIh and ice XIc.
For details on the structure, see \cite[Figure 4]{ref:quantumiceonehstruc}
and also Figure \ref{fig:iceleven} for a caricature.

What is distinctive about ice XI is that it is proton-ordered.
This form of ice is believed to be the most stable (low-energy) 
form \cite{ref:quantumiceonehstruc}.
The ordering of the hydrogen bonds makes ice XI ferroelectric \cite{ref:fukazawa2006existence},
meaning that neighboring dipoles tend to re-enforce each other.
The ferroelectric effect is the result of the accumulation of charge due
to the alignment of dipoles \cite{lrsBIBjc}.

The transition to ice XI from ice I has been extensively studied,
as summarized in \cite{ref:Howe1988PhDiceXI}.

\section{Some other forms of ice}

In addition to ice I, II, and XI, forms of ice denoted by Roman numerals from III to XXI 
have been identified and studied in a 
laboratory \cite{ref:eisenkauzwater,ref:weakstrongHbondIce,ref:everlasthuntice}, so far.
In 2011, only up to ice XV had been described \cite{ref:polymorphicefiveunresolv},
whereas ice XVI was described in 2014 \cite{ref:iceXVI},
ice XVII was described in 2016 \cite{iceXVII}, 
ice XVIII was described in 2019 \cite{millot2019nanosecond},
ice XIX was described in 2021 \cite{ref:iceXIXrelatedtoIceVI,ref:everlasthuntice},
ice XX was described in 2021 \cite{ref:iceXXsuperionic}, and
ice XXI was described in 2025 \cite{ref:iceXXI}.

There are various relations among the various crystalline forms of ice, such as one form 
being a proton ordered form of another, but with the same crystalline structure.
The proton orders can be of two types, one in which the neighboring dipoles tend
to cancel, as in ice II, and the other in which they reinforce, yielding a
ferroelectric material \cite{fatuzzo1967ferroelectricity,
fridkin2013ferroelectricity,jaynesbookferroelectricity,jaynes1953ferroelectricity}.

Ferroelectric properties of water are being explored in protein biophysics
\cite{ref:ferroElectricHydrationShell} and other applications
\cite{ref:oneDwaterFerroElectric}.

\begin{figure}
\centerline{\includegraphics[width=3.5in,angle=-0]{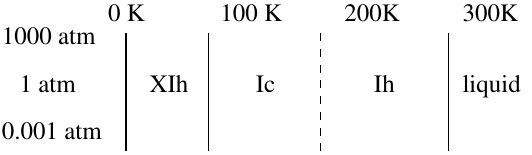}}
\caption{Caricature of the phase diagram for solid and liquid water 
between $10^{-3}$ and $10^3$ atmospheres (atm) and 0 and 300 degrees Kelvin.
Only ice I and ice XI exist in this region of phase space.
The transition to XI is discussed at length in \cite{ref:Howe1988PhDiceXI}.}
\label{fig:lopresspd}
\end{figure}

Figure \ref{fig:lopresspd} gives a caricature of the phase diagram covering
terrestrial pressures and temperatures, and more.
In the range between $10^{-3}$ and $10^3$
atmospheres (atm) and 0 and 300 degrees Kelvin,
it is very simple, involving only ice I and XI.
The dashed line indicates that there is not a clear demarcation of Ic from Ih.
Indeed, it is now suggested \cite{ref:icezeroambiguity} that both phases coexist
at the same temperature and are catalyzed by a new phase dubbed Ice 0.
Moreover, a mixture of Ic layers and Ih layers can form what is called stacking-disordered
ice \cite{ref:stackdisordericeI}.

\begin{figure}
\centerline{\includegraphics[width=3.9in,angle=-0]{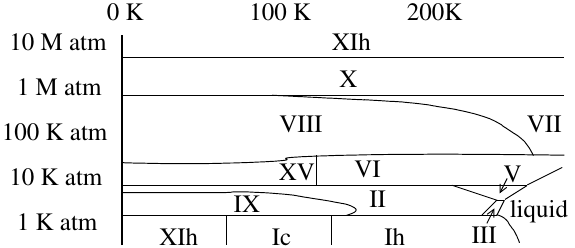}}
\caption{Caricature of the phase diagram for various types of ice and liquid water 
above $10^3$ atmospheres (atm) and between 0 and 290 degrees Kelvin.}
\label{fig:hipresspd}
\end{figure}

Figure \ref{fig:hipresspd} gives a caricature of the phase diagram covering
higher pressures and involves many more forms of ice.

\subsection{Ice III and IX}

The structure of ice III is described in \cite{ref:iceIIIstructureMcFarlan}
and depicted in \cite[Figure 5(a)]{ref:weakstrongHbondIce}.
For further information, see \cite{ref:iceIIIstructureKambPrakash,ref:polymorphicefiveunresolv}.
Ice IX \cite{la1973nearly} is a proton ordered version of 
ice III \cite{ref:iceIIIstructureKambPrakash,londono1993neutron},
depicted in \cite[Figure 5(b)]{ref:weakstrongHbondIce}.
The ordering of protons in ice IX is believed to be antiferroelectric \cite{whalley1968ice},
meaning that neighboring dipoles tend to cancel each other out.

\subsection{Ice IV}

The structure of ice IV is described in \cite{ref:structuriceIV}.
It has a complicated structure involving interpenetrating rings, as shown
stereoscopically in \cite[Figure 3]{ref:structuriceIV}.
Ice IV does not have a stable region in the phase diagram of water, but it 
``can form metastably within the stability field of ice V'' \cite{ref:structuriceIV}.

\subsection{Ice V}

The structure of ice V is described in \cite{ref:structuriceV}.
Its structure unit cell has 28 water molecules.
The tetrahedral structure of water hydrogen bonds is deformed, but not so much
that bifurcated hydrogen bonds are formed \cite{ref:structuriceV}.

\subsection{Ice VI and XV}

The structure of ice VI is depicted in \cite[Figure 6]{ref:weakstrongHbondIce}.
In \cite{ref:weakstrongHbondIce}, the structure is described as
``two identical interpenetrating water frameworks that represent structural
analogues of ... the mineral eddingtonite.''
Ice XV is a proton-ordered version of ice VI;
it is antiferroelectric \cite{salzmann2009ice}.

\subsection{Ice VII and VIII}

The structures of ice VII and VIII are described in \cite{ref:weakstrongHbondIce}
as ``two interposed frameworks of ice Ic,'' with ice VII being proton disordered
and ice VIII being proton ordered.
It appears that the ordering is antiferroelectric \cite{pruzan1993stability}.
Also see \cite{kuo2004structure}.

\subsection{Ice X: symmetric ice}

Ice X departs from our rule regarding hydrogen bonding of water molecules in ice.
Instead, the pressure is so high that the hydrogens are equidistant between oxygens,
meaning the hydrogens are no longer identifiable as being uniquely associated to a
given oxygen \cite{benoit1996new}.

\subsection{Ice XII}

Ice XII is ``a mixture of five- and seven-membered rings of water 
molecules'' \cite{lobban1998structure}.
It is proton disordered \cite{ref:disorderedIceXII}.

\subsection{Ice XIII and XIV}

The structures of ice XIII and XIV are depicted in \cite[Figure 3]{ref:disorderedIceXII}.

\input residualentropy.tex

\section*{Acknowledgments}

We thank Laci Babai and Eric Cances for helpful discussions.

%% file: residualentropy.tex
\section{The residual entropy of water ice}

The concept of residual entropy of ice imples that water molecules can rotate
even at zero degrees Kelvin.
And at higher temperatures, a rotation is even more likely, due to thermal fluctuations.
The rotation of water molecules plays a role in the understanding of the
high dielectric constant of water ice \cite[page 387]{ref:bjerrumicerevuscience}.

When a water molecule rotates in an ice crystal, the ice rules may get violated.
One edge will have two hydrogens on (or near) it, and one edge will have none.
To recover the ice rules, neighboring waters may have to rotate, causing
new violations, and so on \cite[page 388]{ref:bjerrumicerevuscience}.
This process may terminate due to a finite cycle in the graph, or the modifications
may continue indefinitely.

\subsection{Entropy theory}

Physically measurable entropy $S$ was connected to the number $W_N$ of degrees of freedom
in a molecular system by Boltzmann's famous epitaph $S=k\log W$, where $k$ is a
physical constant named for Boltzmann.

By a local argument, Pauling \cite{ref:paulingiceoneh} concluded that
there are $W_N\approx (3/2)^N$ possible orientations of $N$ water molecules.
This leads to a remarkable agreement with the physical measurements
of the residual entropy of water.
However, this argument does not include any nonlocal interactions among
different water orientations.
And at the simplest level of criticism, Pauling's estimate is not an integer.


Pauling's estimate was refined in \cite{ref:NagleResidEntropIce},
based on earlier techniques used in \cite{ref:stillingeresidentropice}.
Without giving all details, the formula \cite[(8)]{ref:NagleResidEntropIce}
says that $W_N=W^N$ where
\begin{equation}\label{eqn:nagleval}
W=\frac32\bigg(1+\sum_n \frac{\phi_n}{3^n}\bigg)\approx 1.5077,
\end{equation}
where $\phi_n$ is the number of cycles of length $n$
that start and end at a fixed vertex, as
given in Table \ref{tabl:nagle} from \cite[Table II]{ref:NagleResidEntropIce}.
Nagle's result \eqref{eqn:nagleval} represents only a half percent change
from Pauling's result.

In an infinite lattice, the numbers $\phi_i$ of such cycles is the same for all vertices.
For a finite lattice, there would be boundary effects.
These effects are ignored in the physics arguments, even though the
basic assumption is that the lattice is finite.

\begin{table}
\begin{center}
\begin{tabular}{|c|c|c|c|c|c|}
\hline
    $n$    & 6 & 8 & 10 &  12 &   14 \\
  $\phi_n$ & 2 & 3 & 36 & 111 & 1068 \\
\hline
\end{tabular}
\end{center}
\vspace{-5mm}
\caption{Values of $\phi_n$ from  \cite[(8)]{ref:NagleResidEntropIce} for ice Ic.}
\label{tabl:nagle}
\end{table}

The residual entropy of water ice is a topic that has been of interest for many decades.
The paper by Lieb \cite{ref:LiebSquareIce} presents a good history up to that time.
Pauling's explanation \cite{ref:paulingiceoneh} gained favor although there had been other 
attempts including quantum-mechanical effects (para versus ortho water)
\cite[page 1148]{ref:heatcapideGiauqStout}; cf.~\cite{ref:coincidenceparaorthowater}.

\subsection{Entropy computation using experiments}

The evidence for the residual entropy $S_0$ was based on heat capacity $C_p$ data as found
in \cite[Table 1]{ref:heatcapideGiauqStout} from 1936.
The formula for computing entropy is based on the thermodynamic
\cite[page 278, line 5]{ref:thermodynformulas} relation
$$
S(T)=S_0+\int_0^T \frac{C_p(t)}{t}\,dt.
$$
If the entropy $S(T)$ at a given temperature $T$ is known, we can write
$$
S_0=S(T)-\int_0^T \frac{C_p(t)}{t}\,dt.
$$
The entropy of ice at the melting point is approximately $41 J/K\cdot mol $.

Computational methods based on classical models of water have been used to estimate
the residual entropy of water ice \cite{ref:resicemulticanpn,ref:entropymdcalc}.
In \cite{ref:resicemulticanpn}, the authors ``pose it as a challenge to experimentalists
to improve on the accuracy of a 1936 measurement'' in \cite{ref:heatcapideGiauqStout}.
Indeed, the early experiments \cite{ref:residentropywater} were done before the
form XI of ice was known.
Any measurements of water ice below 70 degrees Kelvin must be reinterpreted in
this light.

More recent research \cite{ref:dopediceXIentropy,ref:koddopediceXIent} presents data that suggests
a phase change near 72 degrees Kelvin, involving a spike in heat capacity at that temperature.
This data is based on experiments involving ice XI created by doping with various molecules
at a low molar fraction.
The data in \cite[Table 1]{ref:heatcapideGiauqStout} does not include temperatures 
near this phase change.
Given the difficulty of creating ice XI, it may be that the data 
in \cite[Table 1]{ref:heatcapideGiauqStout} did not involve ice XI.
Indeed \cite[section 3.4]{ref:dopediceXIentropy} suggests that
ice Ih ``does not undergo a phase transition by itself.
Only addition of KOH (or RbOH) produces one. The
magnitude of the entropy change of the transition
depends on the concentration of the dopant.''

On the other hand, the data in  \cite{ref:dopediceXIentropy} predicts a residual entropy
of about one third of the Pauling value, not zero as might be expected for pure ice XI,
that is, proton ordered ice in the Ih lattice.
This discrepancy may be due to the dopants used or some as yet undiscovered phase change,
but in any case it suggests that proton ordering may lead to a reduction in the
residual entropy.
In principal, ``proton-ordered forms have no residual entropy'' \cite[page 4]{ref:liquidwaterice}.

\subsection{Zooming out}

There remain questions \cite{ref:replyFerroelectriceXI} about the structure of ice XI,
including whether it is proton ordered or not.
Yet in \cite{ref:plausibleIceXI} it is said that ice XI ``is of interest because
it represents a lower energy state than
the disordered ordinary Ice Ih, and because small domains of Ice XI
have been observed in liquid water \cite{ref:icedynamicsprotonorder}.''
As a result, it is to be expected that ice XI would be the common form of ice
found in space \cite{ref:fukazawa2006existence}.
But ice XI is still difficult experimentally as expressed in \cite{ref:icedynamicsprotonorder}:
``The ground state of one of the most abundant solids in the universe remains
inaccessible in laboratories because its full natural formation can
take up to 100,000 years,'' referencing \cite{fletcher2009chemical}.